\documentstyle{mn}

\newif\ifAMStwofonts

\ifoldfss
  \ifCUPmtlplainloaded \else
    \NewTextAlphabet{textbfit} {cmbxti10} {}
    \NewTextAlphabet{textbfss} {cmssbx10} {}
    \NewMathAlphabet{mathbfit} {cmbxti10} {} 
    \NewMathAlphabet{mathbfss} {cmssbx10} {} 
  \fi
  \ifAMStwofonts
    \ifCUPmtlplainloaded \else
      \NewSymbolFont{upmath} {eurm10}
      \NewSymbolFont{AMSa} {msam10}
      \NewMathSymbol{\upi}     {0}{upmath}{19}
      \NewMathSymbol{\umu}     {0}{upmath}{16}
      \NewMathSymbol{\upartial}{0}{upmath}{40}
      \NewMathSymbol{\leqslant}{3}{AMSa}{36}
      \NewMathSymbol{\geqslant}{3}{AMSa}{3E}

      \let\leq=\leqslant 
      \let\geq=\geqslant 
    \fi
  \fi
\fi 

\ifnfssone
  \newmathalphabet{\mathit}
  \addtoversion{normal}{\mathit}{cmr}{m}{it}
  \addtoversion{bold}{\mathit}{cmr}{bx}{it}
  \newmathalphabet{\mathbfit} 
  \addtoversion{normal}{\mathbfit}{cmr}{bx}{it}
  \addtoversion{bold}{\mathbfit}{cmr}{bx}{it}
  \newmathalphabet{\mathbfss} 
  \addtoversion{normal}{\mathbfss}{cmss}{bx}{n}
  \addtoversion{bold}{\mathbfss}{cmss}{bx}{n}
  \ifAMStwofonts
    \ifCUPmtlplainloaded \else
      \UseAMStwoboldmath
      \makeatletter
      \new@mathgroup\upmath@group
      \define@mathgroup\mv@normal\upmath@group{eur}{m}{n}
      \define@mathgroup\mv@bold\upmath@group{eur}{b}{n}
      \edef\UPM{\hexnumber\upmath@group}
      \new@mathgroup\amsa@group
      \define@mathgroup\mv@normal\amsa@group{msa}{m}{n}
      \define@mathgroup\mv@bold\amsa@group{msa}{m}{n}
      \edef\AMSa{\hexnumber\amsa@group}
      \makeatother
      \mathchardef\upi="0\UPM19
      \mathchardef\umu="0\UPM16
      \mathchardef\upartial="0\UPM40
      \mathchardef\leqslant="3\AMSa36
      \mathchardef\geqslant="3\AMSa3E

      \let\leq=\leqslant 
      \let\geq=\geqslant 
    \fi
  \fi
\fi 

\ifnfsstwo
  \DeclareMathAlphabet{\mathbfit}{OT1}{cmr}{bx}{it}
  \SetMathAlphabet\mathbfit{bold}{OT1}{cmr}{bx}{it}
  \DeclareMathAlphabet{\mathbfss}{OT1}{cmss}{bx}{n}
  \SetMathAlphabet\mathbfss{bold}{OT1}{cmss}{bx}{n}
  \ifAMStwofonts
    \ifCUPmtlplainloaded \else
      \DeclareSymbolFont{UPM}{U}{eur}{m}{n}
      \SetSymbolFont{UPM}{bold}{U}{eur}{b}{n}
      \DeclareSymbolFont{AMSa}{U}{msa}{m}{n}
      \DeclareMathSymbol{\upi}{0}{UPM}{"19}
      \DeclareMathSymbol{\umu}{0}{UPM}{"16}
      \DeclareMathSymbol{\upartial}{0}{UPM}{"40}
      \DeclareMathSymbol{\leqslant}{3}{AMSa}{"36}
      \DeclareMathSymbol{\geqslant}{3}{AMSa}{"3E}

      \let\leq=\leqslant 
      \let\geq=\geqslant 
    \fi
  \fi
\fi 

\ifCUPmtlplainloaded \else
  \ifAMStwofonts \else 
    \def\upi{\pi}
    \def\umu{\mu}
    \def\upartial{\partial}
  \fi
\fi

\def\lsimeq
{\hbox{\raise0.5ex\hbox{$<\lower1.06ex\hbox{$\kern-1.07em{\sim}$}$}}}

\title[ROSAT spectra of EMSS AGN ]
{ The X-ray spectral properties of X-ray selected AGN : ROSAT spectra of 
EMSS AGN}
\author[P.Ciliegi and T. Maccacaro]
       {P.Ciliegi,$^{1,2}$ and T. Maccacaro,$^2$ \\
1. Istituto di Radioastronomia del CNR - Via P. Gobetti 101, 40129 Bologna, 
Italy \\
2. Osservatorio Astronomico di Brera, Via Brera 28, 20121 Milano, Italy\\ }

\date{Accepted May 1996. Submitted December 1995}

\pubyear{1996}

\begin{document}

\maketitle

\begin{abstract}

Using a sample of 63 AGNs extracted from the $Einstein$ 
Extended Medium Sensitivity Survey (EMSS), we study the X-ray spectral 
properties of X-ray selected AGN  in the 0.1$-$2.4 keV ROSAT band.
These objects are all the EMSS AGN detected with more than 300 net counts
in ROSAT PSPC images available from the  public archive (as of May 31, 1995).
A Kolmogorov-Smirnov test
on the redshift and luminosity distributions 
shows that this subsample is representative of the 
whole EMSS sample. 

For the 21 sources detected with less than 600 net counts we characterize 
the spectrum only with the hardness ratio technique. For the other 42 sources
we obtain a detailed spectral analysis fitting the data with two 
different power-law models:  one with N$_H$ fixed at the Galactic value and 
one with N$_H$ as a free parameter. Eight sources ($\sim$ 20 per cent) show a 
significant deviation from a power-law absorbed by Galactic N$_H$,  indicating
soft excess (five sources) or excess absorption (three sources). 
These eight 
sources are analyzed and discussed separately and are excluded from the 
sample used to study the mean X-ray spectral 
properties of the EMSS sources. 

A Maximum-Likelihood analysis is used to find the mean power-law spectral 
index $<\alpha_p>$ and the intrinsic dispersion $\sigma_p$. We find
$<\alpha_p>$=1.42 with $\sigma_p$=0.44. This value is significantly steeper 
($\Delta \alpha \sim$0.4) than the mean $Einstein$/IPC spectral index 
obtained applying the ML analysis on the whole sample of EMSS AGN.  
This result shows that the soft excess already noted in optically selected 
AGN is present also in X-ray selected AGN. The relatively high value obtained
for the intrinsic dispersion confirms that in the soft band AGN are 
characterized by a variety of spectral indices and the increase with respect to the
results obtained from the analysis of Einstein data ($\Delta \sigma_p \sim$0.16)
suggests a further broadening
of the spectral index distribution as one moves to softer energies. 
A comparison between the mean 
spectral index of Radio-quiet and Radio-loud subsamples shows that the mean 
index of the RL sample is flatter than that of RQ, both in the IPC 
($\Delta \alpha \sim$0.3) and in the PSPC ($\Delta \alpha \sim$0.4) data.
This suggests that the additional X-ray component in RL AGN dominates the 
X-ray emission of RL AGN over almost two decades of energy ($\sim$0.1$-$10 keV). 
Finally, we find no significant correlation between the spectral index 
$\alpha_x$ and other physical parameters such redshift, optical and X-ray 
luminosity.

\end{abstract}


\section{INTRODUCTION}
The X-ray spectra of AGN have been studied extensively over the last decade 
using a variety of instruments aboard different satellites (see, among others,
Petre et al. 1984, Reichert et al. 1985,  
Wilkes and Elvis 1987, Canizares and White 1989, Turner and Pounds 1989,
Comastri et al. 1992, Williams
et al. 1992, see also Malaguti et al. (1994) for a catalog of all the 
X-ray spectra of AGN published from the early 70's to the end of 1992). 

These studies indicate that 
the X-ray emission above 1-2 keV is well described by a power-law with a spectral slope 
$\alpha_x\sim$0.5 ($f_{\nu}\propto\nu^{-\alpha}$) for radio-loud (RL) objects and 
$\alpha_x\sim$1.0 for radio-quiet (RQ) objects. Below 1 keV an excess emission 
relative to the flux predicted by an extrapolation of the hard X-ray power-law  is 
observed (Wilkes and Elvis 1987, Comastri et al. 1992). 
Most of the objects studied are among the brightest in the X-ray band 
($f_x\geq1\times10^{-12}$ erg cm$^{-2}$ s$^{-1}$  
0.3$-$3.5 keV) and do not form a complete sample. 
Although the $Einstein$ satellite has produced 
large samples of X-ray selected AGN, the information available
on their X-ray spectral properties is limited by the poor energy resolution 
of the IPC and by the limited statistics of the detected sources.
For the 
{\it Einstein} Extended Medium Sensitivity Survey AGN sample 
(EMSS: Gioia et al. 1990,
Stocke et al. 1991, Maccacaro et al. 1994), for instance,
only the few brightest sources ($e.g.$ MKN 766, MKN 205, 
MKN 1310, PG1416-12, PG1426+015, MKN 877) could be analyzed in some
detail (see Halpern 1981, Elvis et al. 1986, Wilkes and Elvis 1987,
Kruper et al. 1990). 
Since the majority of the EMSS sources had less than 150 
net counts only a statistical analysis of the {\it average} spectral
properties was possible, by means of the analysis of their Hardness Ratio (HR).

Using the HR technique, Maccacaro et al. 
(1988) reported an average energy spectral index $<\alpha_x>$=1.03 with an intrinsic 
dispersion $\sigma$=0.36 for the EMSS AGN sample.   

The better sensitivity, energy resolution (E/$\Delta$E=2.4(E/1 keV)$^{1/2}$ FWHM) and 
spatial resolution of the PSPC detector aboard the ROSAT satellite (Tr\"{u}mper, 1983)
allow us to improve upon the previous study and to extend the spectral analysis
of representative samples of faint AGN ($f_x\simeq1\times10^{-13}$ 
erg cm$^{-2}$ s$^{-1}$) to slightly softer energies.

Recently, Ciliegi et al. (1996)
have reported a detailed X-ray spectral analysis in the 0.1$-$2.4 keV band of a 
complete sample of X-ray selected AGN. Using the 80 AGN (68 QSO and 12 narrow
emission line galaxies, NLXGs) in the Cambridge-Cambridge  ROSAT Serendipity Survey
(CRSS, Boyle et al. 1995) they found that a single power-law plus Galactic absorption 
yields a good representation of the X-ray spectra for almost all the 
sources, 
with a mean energy spectral index $<\alpha_x>=1.32$ (dispersion  $\sigma=0.33$) 
for the quasar 
sample and $<\alpha_x>=1.30$ ($\sigma=0.49$) for the NLXG sample.   
Puchnarewicz et al. (1996) have also analyzed the spectral properties of 
X-ray selected AGN using a sample of 108 objects in the ROSAT 
International X-ray/Optical Survey (RIXOS). They found an average spectral 
index $<\alpha_x>$=1.07 ($\sigma=0.63$),  
marginally flatter than the CRSS average spectral index. 
Finally Laor et al. (1994) and  Walter and Fink (1993) have analyzed optically 
selected samples of quasar obtaining a mean ROSAT  spectral index of 
$\alpha\simeq$1.50.

In this paper we report the soft X-ray spectral analysis of all the 
EMSS AGN detected with more than 300 net counts in the ROSAT/PSPC public (as of 
May 31, 1995) observations. In section 2 we define the sub-sample of the EMSS AGN 
used, while in section 3 we describe the method of analysis.  
We report and discuss the results in section 4 and 5 respectively, while  
in section 6 the conclusions and a summary are presented.
Throughout the paper a Hubble constant of $H_0=50\,$km$\,$s$^{-1}\,$Mpc$^{-1}$
and a deceleration parameter $q_0=0$ are assumed.

\section{THE SAMPLE}
We searched in the ROSAT archive for ROSAT (PSPC) images 
containing pointed or ``serendipitous'' observations of EMSS AGN.
The first selection criterion used is an off-axis angle $\theta$ (the
distance between the source and the center of the PSPC field) smaller
than 50 arcmin. We find data for 203 different EMSS AGN.
Because our aim is a detailed study of the X-ray spectral properties,
we then selected only those 
sources detected with more than 300 measured net counts. 
For sources observed more than once, we retain only the observation
where the object is detected with the highest number of net counts.
We find 63 EMSS AGN satisfying the above criteria; they are listed in Table 1
which is organized as follows: source name, followed by
redshift, the two 
point spectral index between radio and optical band ($\alpha_{ro}$, 
from Stocke et al. 1991), 
the extraction radius $r$ (arcmin) calculated as described in section 3, the 
off-axis angle $\theta$ (arcmin), the net counts detected in the 0.1$-$2.4 keV band
within the $r$ radius circle  with the relative error computed as the
square root of the total (source + background) observed counts,
and the sequence number of the ROSAT observations. Sources for which
``normalized" net counts  have been computed (see Section 3 for details) are
indicated with an asterisk.

Five of these 63 objects are radio-loud objects on the basis of the standard 
radio to optical spectral index criterion ($\alpha_{ro}\geq$0.35). 

A Kolmogorov-Smirnov (KS) test shows that the redshift, X-ray and optical
luminosity distributions of the 63 EMSS AGN discussed in the present paper
are not significantly 
different (at a confidence level greater than 90 per cent) from the 
distributions of the whole EMSS AGN sample . Moreover we 
have checked whether the requirement that the sources have more than 300 
net counts could introduce a bias in the spectral index distribution due to the
fact that steeper spectrum sources could have higher ROSAT count rates. 
We have searched for a possible correlation between net counts and spectral 
indices but we found no evidence for it. 

We are therefore confident that, although in our 
analysis we used only $\sim$ 15 per cent of the EMSS AGN,
the results obtained are representative of the whole sample. 
We can then compare the mean ROSAT/PSPC spectral properties of
the subsample of 63 objects with the mean $Einstein$/IPC spectral 
properties of the whole sample, taking advantage of the large size of the latter. 

\section{DATA ANALYSIS}
For each source we fit the X-ray image with a two-dimensional gaussian
to determine its centroid. Subsequently we extract the total counts
in the $0.1-2.4\,$keV band using a circle centered on the source
centroid. To optimize the signal to noise ratio the extraction radius $r$ is
chosen to be the distance from the centroid
at which the radial profile of the point source meets the background
level. All
the extraction radii lay in the range
$2.0\,^{\prime}\leq r\leq5.0\,^{\prime}$. The background counts
are estimated in an
annulus centered on the source with inner radius $r+1\,^{\prime}$ and
outer radius $r+3\,^{\prime}$.
When other sources are detected within the 
background region, they are removed using a cutting circle with a radius
chosen to contain
at least 95 per cent of the source counts (Turner and George 1992).
For the 11 sources detected near other sources or near the window support 
structure, an azimuthal sector centered on the source position is 
excluded from the circle used for the counts extraction (and 
from the annulus used for the background determination) to avoid 
contamination or shadowing. 
The net counts thus obtained
are used for the spectral analysis. The X-ray 
flux of these sources, however, is determined normalizing the net counts to a full 
circle, assuming that the source net counts are azimuthally uniformly distributed. 
Thus, for example, when we exclude a 90$^{\circ}$ azimuthal sector, 
the normalized net counts (and relative error) are obtained 
multiplying the measured counts  (and error)  by 360/(360-90). 
For the four sources (MS0439.7$-$4319, MS0919.3+5133, MS1059.0+7302, 
MS1408.1+2617) detected near the window support structure we included 
a systematic error of 10\% in the X-ray flux by adding it in quadrature
to the statistical error. 
Sources for which normalized net counts have been computed are indicated
in Table 1.

For all sources we have computed the energy
spectral index $\alpha_x$ of the power-law using the hardness ratio HR and fixing
N$_H$ at the Galactic value.  The hardness ratio HR is defined as HR = (H$-$S)/(H+S)
where S is the number of net counts detected in the 
0.11$-$0.43 
keV band and H is the number of net counts detected in the  
0.5$-$2.02 keV band (see Appendix A in Ciliegi et al. 1996 for a full 
description 
of the ROSAT hardness ratio technique). 
Following Hasinger  (1992), before computing the hardness ratio, we have 
corrected the net counts of each source for the energy dependence of 
the Point Spread Function (PSF). 
Of the 63 sources, 42 are detected with more than 600 net counts. For these
sources a more detailed spectral analysis is possible. We have thus considered 
31 of the 34 
energy channels from the MPE Standard Analysis
Software System (SASS)  pipeline processing excluding the first
two channels ($<0.1\,$keV) and the last channel ($>2.4\,$keV)
because the response of the instrument is
not well-defined at the extremes of the energy range. The
spectrum is then binned  to obtain at least 20 net counts and 
a signal to noise ratio $S/N\geq 3$
in each bin so that the $\chi^2$ statistic could be applied.
Finally we make two different power-law
fits: (1) with N$_H$ fixed at the Galactic value and (2) with N$_H$ as free parameter. 
Model fits are 
carried out using the XSPEC software package, and the best fit model
parameters are 
obtained by $\chi^2$ minimization. Following Fiore et al. (1994), we have used
the 1992 March response matrix for observations made before October
1991 and the 1993 January
response matrix for observations made after October 1991. 

\section{RESULTS}
The results of the spectral fits are given in Table 2.  
For each source we report the results of the hardness ratio analysis (columns 2 and 3) 
and of the detailed spectral analysis (columns 4 - 9). 
The table is organized as follows:

{\it Column 1:} Source name. Radio-loud AGN are indicated with ``R''.

{\it Column 2:} Hardness Ratio ($HR$).

{\it Column 3:} Energy spectral index $\alpha_{HR}$ obtained from the 
$HR$, fixing  $N_H$ =  $N_{H_{\rm Gal}}$. 

{\it Column 4:} Best fitting spectral slope $\alpha_x$ for fit 1
 (first row) and for fit 2 (second row). 

{\it Column 5:} $N_{H_{\rm Gal}}$ used for fit 1 (first row) and 
the best fitting  $N_H$ for fit 2 (second row). 

{\it Column 6:} $\chi^2$ of the fit ($\chi^2_{\rm FIT}$) and 
  the number of degrees of freedom (dof) for fit 1 (first row) and 
fit 2 (second row).

{\it Column 7:} Probability for $\chi^2\geq\chi^2_{\rm FIT}$.

{\it Column 8:} F-test probability (Bevington and Robinson 1992)
that the reduction  in $\chi^2_{\rm FIT}$ with the addition 
of  $N_H$ as a free parameter is not statistically significant ($P(F>F_{\rm FIT}$)).

{\it Column 9:} X-ray flux in the 0.1$-$2.4 keV band obtained using the
fit 1 parameters  or, for sources with less than 600 net counts, 
using $\alpha_{HR}$ and Galactic $N_H$. Flux errors are computed 
from photon-counting statistics only ($i.e.$ without taking into account 
the error on the spectral slope and on N$_H$). 

The errors represent the 68 per cent confidence intervals for a single 
interesting parameter when N$_H$ is fixed, and for two interesting 
parameters when both the X-ray energy spectral index $\alpha_x$ and N$_H$ 
are free to vary. 

A comparison of fit 1 and 2 allows us to determine whether
there is evidence for a significant intrinsic excess absorption or
excess emission 
relative to a single power law model plus Galactic absorption.
As shown in Table 2, in eight of the 42 objects for which we
have enough net counts to allow a spectral fit, there are significant 
deviations from a simple power-law plus Galactic absorption 
(i.e. $P(F>F_{\rm FIT})<0.01$).
These eight sources, that will be separately discussed in $\S$ 5.1, represent about 
20 per cent of the sources with more than 600 net counts. 

Finally, we have used these 42 sources to test the reliability of the hardness ratio 
technique by 
comparing the spectral indices $\alpha_{HR}$ derived from
the hardness ratio with those derived from the detailed spectral
analysis (fit 1: $N_H$ fixed
at the Galactic value). The result is illustrated in Figure 1.  The 
slope derived by the two method are very well correlated through a wide 
range of values ($\alpha_x \sim 0.4 - 2.2$). For $\alpha_x>$ 2.2 
(3 objects) the slopes 
obtained from the hardness ratio are lower ($\Delta \alpha \sim$0.2) 
than those obtained from the detailed spectral analysis. 
This same difference was evident also in the results of the X-ray 
spectral analysis of the CRSS AGN sample (Ciliegi et al. 1996). 
Since all the EMSS AGNs  analyzed solely with the HR 
show an energy spectral index $\alpha_x<$2.1
(see Table 2) we are confident that can safely combine the results 
obtained with the two techniques. 
 Of course one should keep in mind that the hardness ratio 
technique is not able to recognize those sources showing a significant 
deviation from a power-law model plus Galactic absorption, since its application
rests on the assumption of the knowledge of the spectral form.
However, on the basis of the results 
obtained from the detailed analysis of the 42 sources detected with more than 
600 net counts 
we expect that this assumption ($i.e.$ X-ray spectra well fitted by
a simple power-law plus Galactic absorption) may not be appropriate only for 
$\sim$20 per cent of the 21 sources analyzed with this technique
($i.e.$ $\sim$ 4 sources). 

We have therefore removed the 8 sources that show significant deviations from
a single power-law model plus Galactic absorption to obtain a sample of 55
AGN which we use in the following.
Given the small number of objects for which the HR results may be unreliable
($<$8 per cent 
 of the whole sample of 55 objects) the results obtained should not 
be strongly affected by their possible inclusion.

\section{DISCUSSION}
\subsection{\it 5.1 Sources with a significant deviation from a simple 
power-law plus Galactic absorption}
As discussed above, eight sources show a significant deviation from a simple 
power-law fit plus Galactic absorption. 
 Five of these  sources show a best fit value of N$_H$ 
lower than the Galactic value, indicating an excess of soft emission relative 
to a single 
power-law fit with N$_H$=N$_{H_{\rm Gal}}$, while the other three sources show 
excess absorption  (N$_H>$N$_{H_{\rm Gal}}$). For only two of these eight sources 
(MS1215.9+3005 and MS1747.2+6837) there are ROSAT/PSPC spectra 
already published in the literature (Molendi and Maccacaro 1994, Brandt et 
al. 1994). For the other six sources we plot in Figure 2 a two dimensional 
contour plot for a single power-law model with spectral slope $\alpha_x$
and equivalent hydrogen absorbing column density N$_H$. 
Below we briefly discuss each of these sources in turn. 

\vspace{2mm} 

{\bf MS0104.2+3153} This X-ray sources is a QSO-Galaxy pair discovered as a 
serendipitous source in the EMSS. Based upon the IPC data (52 net counts, 
Gioia et al. 1990) the true identity of MS0104.2+3153 was 
ambiguous due to the presence in the error circle ($\sim40^{\prime\prime}$)  
of a radio-quiet broad absorption line (BAL) QSO (z=2.027) only 
10$^{\prime\prime}$ away from a giant elliptical galaxy (z=0.111) at the 
center of a compact group of galaxies (Stocke et al. 1984). 
Successively, this source was observed with the Channel Multiplier Array
(CMA) aboard the EXOSAT satellite (Gioia et al. 1986). Since no X-ray 
source was apparent in the deep (160 ksec) EXOSAT observation, 
Gioia et al. (1986) concluded that the X-ray source is variable and thus 
that the QSO is the strongest candidate for the 
optical identification of MS0104.2+3153 excluding any contribution to the 
X-ray flux from the giant elliptical galaxy or from the cluster of 
galaxy. Moreover, Gioia et al. (1986) concluded that the EXOSAT CMA data 
implied a substantial decrease in the X-ray luminosity of this object, 
unless its soft X-ray spectrum is unusually and extremely flat 
($\alpha_x<$0.2) or this object exhibits a significant amount of intrinsic 
absorption (N$_H>$ 2$\times10^{21}$ cm$^{-2}$; quite possible for a 
BAL QSO). 

The ROSAT observation allow us to study  in more detail the 
X-ray emission from this source. Unfortunately, in spite of the improved 
spatial resolution of the ROSAT/PSPC detector, we are not yet able to 
confirm, on positional ground, the optical identification of this source. 
As shown in Figure 2a (see also Table 2) this source has a best fit 
N$_H$ value larger than the Galactic value at the 99 per cent confidence level. 
We fit the absorption excess modelling the data with a power-law 
plus Galactic absorption plus a second
absorption component. We make two different fits, with the additional 
absorption component fixed at the redshift of the cluster (z=0.111, fit A) 
and at the 
redshift of the QSO (z=2.027, fit B).  
In table 3 we report the results of these 
spectral models. The PSPC cannot constrain the redshift
of the absorbing material, both  models give, in fact, 
a good fit of the data (see Table 3). 

The spectral results (models A e B, Table 3) show that the X-ray spectrum 
of this source is characterized by a very steep spectral index 
($\alpha_x>$ 2.9) and by a strong excess absorption relative to the 
Galactic value (N$_H>4\times10^{21}$). Therefore, the non detection with the 
EXOSAT CMA can not be attributed to an unusually and extremely flat spectrum. 
However the possibility that the non detection is not due to an X-ray flux 
variability but  to a significant amount of intrinsic absorption can
not be excluded.

In conclusion, the ROSAT observation of MS0104.2+3153 does not allow us 
to obtain a definite identification of the X-ray source.

\vspace{2mm} 

{\bf MS0132.5$-$4151}
This source shows a best fit value of N$_H$ lower than the Galactic value at 
the 99 per cent confidence level (see Figure 2b).  We parameterized this soft 
excess fitting the data with a broken power-law  with N$_H$ fixed at the 
Galactic value. We find that the spectral index of the soft power-law 
is poorly constrained while the hard power-law has a spectral index 
$\alpha_x=2.21\pm0.65$ (see Table 3).  A two component model, consisting 
of a thermal component (black body, bremsstrahlung or Raymond-Smith dominating
the soft part of the spectrum) plus a power-law does  not yield a good fit 
of the data (P($\chi^2)<$0.01). 

\vspace{2mm} 

{\bf MS0144.2$-$0055} This source shows excess absorption relative to the 
Galactic value (see Figure 2c). The absorption excess in MS0144.2$-$0055 can 
be well parameterized by fitting the data with a power-law plus Galactic
absorption plus a second  absorption component 
(N$_H$=1.5$^{+0.8}_{-0.7}\times10^{20}$ cm$^{-2}$) 
at the redshift of the source (see Table 3). 

\vspace{2mm} 

{\bf MS0310.4$-$5543} This source was included in the sample of 53 AGNs which 
exhibit ultra-soft X-ray excess (Ultra Soft Survey, USS) selected from 
$Einstein$ IPC sources (Cordova et al. 1992, Puchnarewicz et al. 1992). 
However, given the few net counts detected by $Einstein$ ($\sim$ 40 net 
counts) only a limited spectral analysis was possible. 

The ROSAT observation ($\sim$ 7700 net counts detected) gives the opportunity
to study the X-ray 
spectrum of this source in more detail. 
As shown in Figure 2d,  the ROSAT data show a soft emission excess 
relative to a single power-law with  N$_H$=N$_{H_{\rm Gal}}$. We find that the soft 
excess in this source is well parameterized by fitting the data with a 
broken power-law with N$_H$ fixed at the Galactic value. It gives 
$\alpha_{SOFT}=3.89\pm1.09$, $\alpha_{HARD}=1.54\pm0.13$, 
E$_{break}=0.25\pm0.13$ keV and $\chi^2$/dof=23.13/23 (see Table 3). 

\vspace{2mm} 

{\bf MS0919.9+4543} This source shows excess absorption relative to the 
Galactic value (see Figure 2e). As for MS0144.2$-$0055, also for this source 
the absorption excess can 
be well parameterized by fitting the data with a power-law including a 
Galactic  absorption  plus another absorption component
(N$_H$=2.7$^{+1.6}_{-1.3}\times10^{20}$ cm$^{-2}$) 
at the redshift of the source (see Table 3). 

\vspace{2mm} 

{\bf MS1215.9+3005} This source is the well know Seyfert 1 galaxy MKN 766. 
A detailed 
analysis of the ROSAT observations of this source is reported by 
Molendi, Maccacaro and Schaeidt (1993) and
Molendi and Maccacaro (1994) while the results of the ASCA observation
are reported by Leighly et al (1996). 
Its X-ray spectrum is very complicated, 
therefore we refer to these works 
for a detailed discussion of the X-ray properties of this source. 

\vspace{2mm} 

{\bf MS1747.2+6837} This is the Seyfert 1 galaxy Kaz 163. The ROSAT/PSPC data  
of this source were previously analyzed by Brandt et al. (1994). They find 
that a single power law model plus galactic absorption is not a good 
representation of the data since it would imply an N$_H$ column density 
lower than Galactic value. However they showed that a broken power-law is 
a good representation of the spectrum, yielding
N$_H$=5.01$^{+1.40}_{-1.27}\times10^{20}$ cm$^{-2}$, 
$\alpha_{SOFT}$=2.59$^{+0.81}_{-1.03}$, 
$\alpha_{HARD}$=1.54$^{+0.10}_{-0.10}$ with $\chi^2$/dof=23.8/28. 

In our analysis we confirm the results obtained by Brandt et al. (1994). 
However, since Brandt et al. (1994) showed that there are no indications 
of excess absorption (N$_H\sim$N$_{H_{\rm Gal}}$) in the broken power-law model, 
we parameterized the soft excess fitting the data with a broken power-law 
fixing N$_H$ at the Galactic value. With one less free
parameter in the fit, we can better constrain the slopes of the 
broken power-law. We find $\alpha_{SOFT}=2.93\pm$0.50 and 
$\alpha_{HARD}=1.55\pm$0.09 (see Table 3).
\vspace{2mm} 

{\bf MS2340.9$-$1511} This source shows a best fit value of N$_H$
lower than the Galactic value, although only at the 68 per cent confidence level
(see Figure 2f). 
 Both the two power-law models with N$_H$ fixed at the 
Galactic value and with N$_H$ as a free parameter do not yield a good fit of the 
data (P($\chi^2)\sim1.5 \times 10^{-8}$ for both fits, see Table 2). 
We find that the X-ray spectrum of MS2340.9$-$1511 can be well fitted by a 
two component model consisting of a thermal component (kT=0.52$\pm$0.11 keV) 
based on the model 
calculations of Raymond and Smith (1977) with abundances fixed at the cosmic 
value, plus a power-law component ($\alpha=2.06\pm0.07$).
In table 3 we report the results for the 
Raymond-Smith + power-law fit.  Other spectral models used to fit the 
data (broken power-law, black-body + power-law, bremsstrahlung + power-law) 
do not yield a good description of the data (P($\chi^2)<$0.01).

\subsubsection{Warm Absorber}
For the four sources that show a best fit value of N$_H$ lower than 
Galactic value (MS0132.5$-$ 4151, MS0310.4$-$5543, MS1747.2+6837, 
MS2340.9$-$1511 $-$ MKN766 is not considered here for the reason stated
above) we investigated 
the possibility that the soft excess could be a signature of an 
absorption produced not by neutral gas but by partially ionized ``warm'' 
material. A signature of such material is a flattening of the spectrum above 
the oxygen K-edge at $\sim$ 1 keV, and an excess flux below the edge, at 
$\sim$ 0.25 keV, where the ionized lighter elements become transparent. 
Given the limited spectral resolution of the ROSAT PSPC, warm absorbers could 
led to apparent soft excess in the X-ray spectra. 

Fiore et al. (1996) have considered a number of warm absorber models and 
have computed for different values of the spectral index $\alpha_x$, of 
the absorption column density N$_H$ and of the ionization parameter $U$, the 
expected ratio of PSPC counts: R1 = (1.2$-$2.4 keV)/(0.6$-$1.0 keV) and 
R2 = (0.6$-$1.0 keV)/(0.2$-$0.4 keV). They have shown that ``warm absorbers''
cluster in a well define region of the R1 vs. R2 plane\footnote{Note that 
the label along X-axis of Figure 7 of Fiore et al. (1993) should read 
Counts(0.6$-$1.0) / Counts(0.2$-$0.4) and not Counts(0.2$-$0.4) / 
Counts(0.6$-$1.0) (Fiore, private communication)}.

We have computed the position of the four above mentioned sources in the R1
vs. R2 plane and discovered that all of them fall far away ($>6\sigma$) 
from the ``warm absorbers'' region (see Figure 3). Therefore we can conclude 
that absorption by partially ionized ``warm'' material can be excluded for 
MS0132.5$-$4151, MS0310.4$-$5543, MS1747.2+6837 and MS2340.9$-$1511.

\subsection{Analysis of the power-law spectral index} 
Within the general assumption that both the measurement errors and the 
underlying spectral index distributions can be described by a 
Gaussian, Maccacaro et al. (1988) used the Maximum-Likelihood 
(ML) analysis (see also Worrall and Wilkes 1990) to obtain the 
mean spectral index ($<\alpha_p>$) 
and the intrinsic spread in spectral
slope ($\sigma_p$) of each class of extragalactic objects in the 
EMSS (we use here the same notation of Maccacaro et al. 1988, where 
the subscript "p" refers to the parent population). 
For the AGN sample they found an average spectral index 
$<\alpha_p>$=1.03$^{+0.05}_{-0.06}$ with an intrinsic 
dispersion $\sigma_p$=0.36. 

In Figure 4 we show the distribution of the EMSS AGN energy spectral
indices obtained from our analysis of the ROSAT data. Because a KS test  
shows that the spectral index
distribution of the subsample of EMSS AGN  used 
is not significantly different from a normal distribution 
($>$ 90 per cent confidence level), we used the ML analysis to calculate 
the mean ROSAT spectral index and the intrinsic dispersion. 

Since at the time of the Maccacaro et al. (1988) analysis of the
spectral properties of the EMSS AGN there was still a large number (236)
of unidentified sources (of which 212 have since been identified) we have also
repeated, for a more meaningful comparison, the ML analysis using the 
IPC data on the current (Maccacaro et al. 1994) EMSS AGN sample. 

We have therefore considered 
437 AGN (391 RQ and 46 RL) for the ML analysis, excluding  
26 sources (23 RQ and 3 RL) for which
 the hardness ratio is so extreme to make the slope determination
meaningless.

Table 4 and Figure 5 give the results of 
the ML analysis. In Table 4 we report the mean spectral index 
$<\alpha_p>$  and the intrinsic dispersion 
$\sigma_p$ obtained with the ML analysis and, for comparison, 
the weighted mean $\overline{\alpha}_{\mbox {\tiny WM}}$. 
Figure 5 shows  
the 90 per cent confidence contour for two interesting parameters 
for each data set ($Einstein$/IPC and ROSAT/PSPC).

A comparison of our IPC results on the 
whole EMSS AGN sample with the results obtained by Maccacaro et al. (1988), 
shows that we obtain the same mean spectral index with a small decrease 
in the intrinsic dispersion ($\Delta\sigma_p$=0.08). On the other hand, 
Table 4 and Figure 5 show that the mean PSPC spectral 
index is significantly steeper than the mean IPC spectral index 
($\Delta\alpha\simeq$0.4).

A similar discrepancy 
between IPC and PSPC slopes of AGN has been pointed out also by 
Laor et al. (1994) and Fiore et al. (1994). 
This difference is probably due to the "ultra-soft excess" below $\sim$0.3 keV
first noted by 
Wilkes and Elvis (1987) in the IPC X-ray spectra of quasars
and interpreted as the high-energy tail of the hot thermal component 
dominating the UV emission (the big blue bump). 
Due to the softer ROSAT/PSPC band ($\sim 0.1-2.4$ keV) compared to the 
$Einstein$/IPC band ($\sim 0.3-3.5$ keV) the "ultra-soft excess" dominates 
in the ROSAT spectra and causes a steepening of the X-ray spectra when fitted 
with a single power-law. However, even if this "ultra-soft component" becomes  
dominant in the ROSAT band compared to the other components proposed to explain 
the X-ray emission above 0.5 keV (see Wilkes and Elvis 1987), we expect
that its introduction 
 causes a broadening in the spectral 
index distribution. In fact, because this thermal component comes from the 
inner accretion disk around the central black hole, different physical and/or 
geometrical properties of the latter may be responsible of different values 
of the observed spectral slope, with a consequent broadening of the spectral 
index distribution. 
 The ML analysis shows (see Table 4 and Figure 5) that the 
intrinsic spread in the ROSAT spectral slopes is, in fact, greater than 
the intrinsic spread in the $Einstein$ spectral slopes ($\Delta\sigma_p$=0.16). 
To ensure that this difference is not due to the different 
number of objects  used (411 for IPC data and 55 for PSPC data), we repeated the
ML analysis using only the IPC spectral indices of the 55 EMSS AGN that we used in 
our ROSAT analysis. For these sources we find $<\alpha_p>$=1.14 and $\sigma_p$=0.31.
The ROSAT spectral indices show a greater intrinsic dispersion 
($\Delta\sigma_p$=0.13)
compared to the $Einstein$ spectral indices also for the same data set. 
We therefore suggest that the difference between the ROSAT and $Einstein$ spectral 
indices is due to the presence of the "ultra-soft excess" in the ROSAT/PSPC 
band. However, one should consider the possibility that up to 50\% of 
this difference ({\it i.e.} $\Delta \alpha \lsimeq$0.2) may be due to 
calibration errors in the PSPC and/or IPC instruments. In fact, as suggested 
by Fiore et al. (1994) (see also Turner 1993), the maximum amplitude of the 
systematic error due to calibration is  $\sim0.2$ 
on the spectral index $\alpha_x$.

Finally, considering only the ROSAT data, 
our values of the mean spectral index are consistent with previous values 
obtained for X-ray selected AGN. Using the CRSS sample, Ciliegi et al. (1996) 
find a mean spectral index $<\alpha_x>$=1.32 ($\sigma_{p}$=0.33) for the 
quasar sample and $<\alpha_x>$=1.30 ($\sigma_{p}$=0.49) for the 
NLXG sample, while Puchnarewicz et al. (1996) using the RIXOS AGN sample 
find  $<\alpha_x>$=1.07 ($\sigma$=0.63).  
These mean spectral indices of X-ray selected AGN 
 although marginally flatter, are also consistent with the values obtained 
for optically selected AGN. Laor et al. (1994) and Walter and Fink (1993) 
in fact, using samples of optically selected AGN, find a mean value of the ROSAT/PSPC
spectral index of 1.50 ($\sigma =0.40$ and $\sigma =0.30$ respectively).

\subsection{The radio-loud and the radio-quiet subsample} 
The spectral differences, in the X-ray band, between RL and RQ AGN have 
been studied by many authors. Wilkes and Elvis (1987) showed that RL 
objects have flatter X-ray spectra 
($\alpha_x\sim$0.5 in the 0.3$-$3.5 keV band) compared to RQ
objects ($\alpha_x\sim$1.0). Lawson et al. (1992), using
EXOSAT data, showed that RL objects have X-ray spectral indices
consistent with a single value (i.e. consistent with a dispersion
$\sigma$=0.0), whereas RQ objects show a large spread in indices
($\sigma>$0.10).

Using the ML analysis we obtain the mean spectral index and the intrinsic 
dispersion for the RQ and RL EMSS AGN subsamples using the $Einstein$/IPC 
data (368 RQ and 43 RL) and the ROSAT data (50 RQ). Given the small number
(5) of ROSAT RL AGN that we have analyzed, we do not apply the ML 
analysis to this subsample but we obtain a mean spectral index simply 
with the weighted mean.  Table 4 and Figure 6 give the results of the 
ML analysis. It is clear that the mean index for the RL sample is flatter 
than that for RQ, and further that this group of objects shows no 
intrinsic dispersion, i.e. they are compatible ($>$ 90 per cent
 confidence level) 
with a single spectral index, whereas there is an intrinsic dispersion 
present among the RQ samples (both from $Einstein$ and ROSAT data). 
The same results were obtained by Lawson et al. (1992) with the EXOSAT 
data for a sample of 13 RQ and 18 RL quasars. 

The flatter and possibly unique X-ray spectral index that we find also for 
the EMSS RL AGN with the IPC data, strengthen the scheme (Zamorani et 
al. 1981, Wilkes and Elvis 1987, Shastri 1991, Lawson et al. 1992) 
in which there is an additional component present in radio-loud objects
which produces this single dominant power-law. This could be associated 
with beaming and jets present in many radio-loud quasar, as recently 
discussed by Kembhavi (1993) and Ciliegi et al. (1995).

Finally, Table 4 shows that also in the ROSAT data the RL sample 
shows a flatter spectral index than the RQ sample 
($\Delta\overline{\alpha}_{\mbox {\tiny WM}}=0.42$). It thus seems that the 
additional X-ray component in the RL AGN dominates the X-ray spectra of 
these sources over almost two decades of energy ($\sim0.1-10$ keV).

\subsection{Correlation analysis} 
In this section we describe the search for correlations between $\alpha_x$ and 
other physical parameters, in particular redshift, optical and X-ray
luminosity ($L_{\rm 2500\AA}$ and L$_x$) and 
optical to X-ray luminosity ratio ($L_{\rm 2500\AA}/L_x$).

The  significance of the correlations is tested using the Spearman Rank
correlation coefficient $r_S$ and the relative probability $P_r$ that
an observed correlation could occur by chance for uncorrelated data
sets (see Press et al. 1992).
A correlation is taken to be significant when $P_r\leq 0.01$. A
 summary of all the correlation coefficients and their significance is
given in Table 4, while Figure 7  shows redshift ($z$),
 optical luminosity ($L_{\rm 2500\AA}$), X-ray luminosity ($L_x$)  and
$L_{\rm 2500\AA}/L_x$, respectively,  as a function of the X-ray energy
spectral index $\alpha_x$.

Given the hidden dependence of the normalization at 2 keV on the 
spectral slope $\alpha_x$ obtained from PSPC spectra
(see Ulrich and Molendi 1995 for a detailed discussion of this point),
we use,  in the correlation
analysis the broad band (0.1$-$2.4 keV) X-ray 
luminosity ($L_x$) which is much less sensitive to changes in
$\alpha_x$. For this reason we do not use the usual two point
spectral index $\alpha_{ox}$ but instead we use the ratio between 
the optical (at
2500\AA) and broad band X-ray luminosities ($L_{\rm 2500\AA}/L_x$).

Table 4 shows that no significant correlations are found. The lack 
of correlation between $\alpha_x$ and $z$ for the EMSS AGN suggests that 
for these sources the power-law spectrum in the source rest frame extends 
from the soft ($\sim 0.1-2.4$ keV) into the hard X-ray band ($\sim 0.3-7.0$ 
keV for the highest redshift object). Similar results are obtained for X-ray 
selected quasars by Ciliegi et al. (1996) and by Puchnarewicz et al. (1996) 
and for optically selected quasars by Canizares and White (1989) 
using $Einstein$/IPC data and by Williams et al. (1992) using Ginga data. 

These results appear to be at odd with the result obtained by Schartel et al. 
(1992). Using the RASS data for 162 
strongly-detected optically-selected QSOs Schartel et al. (1992)
find a flattening of the spectral slope  with increasing redshift from  
$\alpha_x\sim 1.5$ to $\alpha_x \sim 0.8$ between $z=0.2$ and $z\sim 2$.
The fact that both the EMSS and CRSS sample
do not show any correlation between spectral index and redshift 
strengthen the suspect  that 
the correlation observed in the Schartel et al. 1992 sample is spurious
and due 
to an increasing fraction of RL AGN (that show flatter spectra) at 
high redshift. In fact, unlike the analysis carried out for the ROSAT 
spectra of the EMSS and CRSS sample, Schartel et al. (1992) did not make 
the important distinction between RL and RQ objects. 
Indeed the Ginga data appear to show 
a flattening of  $\alpha_x$ with redshift but the correlation disappears when RL and 
RQ objects are considered separately (Williams et al. 1992). 

Finally, the lack of a correlation between $\alpha_x$ and optical and 
X-ray luminosities is in agreement with the results obtained by 
Ciliegi et al. (1996) for the CRSS AGN and by Laor et al. (1996) 
for 23 QSOs of the Bright QSO Survey (BQS).

\section{CONCLUSION} 
Using ROSAT/PSPC observations in the public archive (as of May 31, 1995) 
we have analyzed the X-ray spectra
of 63 EMSS AGN. These objects are all the EMSS AGN detected by ROSAT 
with more than 300 net counts. 
A comparison with the whole sample of EMSS AGN
shows that this subsample
is not significantly different (redshift and luminosities distributions)
from the whole EMSS AGN sample. This allows 
us to compare the PSPC results with the mean IPC spectral 
properties of the whole EMSS AGN sample.    
Our major results are the following: \\ 

1. Of the 42 sources with more than 600 net counts for which a detailed 
analysis was possible, eight ($\sim$20 per cent)
show a significant deviation from a single power-law 
model plus Galactic absorption. These eight sources were excluded from 
the sample used to study the mean X-ray spectral properties of the 
EMSS sources. Five of these 
sources show a soft emission excess relative to a single power-law fit with 
N$_H$=N$_{H_{\rm Gal}}$ whereas the other three show a significant absorption 
excess (N$_H>$N$_{H_{\rm Gal}}$). Using a color-color diagram developed 
by Fiore et al. (1993) we can exclude the possibility that the 
soft emission excess is due to absorption by partially ionized 
``warm'' material.   

2. The mean ROSAT/PSPC energy spectral index obtained with the 
Maximum-Likelihood analysis 
is $<\alpha_p>$=1.42 with an intrinsic dispersion $\sigma_p$=0.44. This value 
is in agreement with the ROSAT spectral indices obtained for other X-ray and
optically selected AGN, and
is significantly steeper ($\Delta\alpha\simeq$0.4) than the mean $Einstein$/IPC 
spectral index obtained on the whole sample
of EMSS AGN. Moreover we find a significant increase in the intrinsic 
dispersion of the ROSAT spectral indices compared to the $Einstein$ data 
($\Delta\sigma_p$=0.16). The steepening of the ROSAT mean spectral index is 
probably due to the fact that in the ROSAT band (softer than the IPC band) 
the "ultra-soft excess", already noted in the 
spectra of many AGN below $\sim$0.3 keV, becomes dominant. 
In this scheme, the broadening of 
the ROSAT spectral index distribution can be explained considering that the 
slope of the ultra-soft component may differ from object to object, reflecting
different physical and/or geometrical properties of the inner accretion disk 
with which it is associated. 

3. The ML analysis of the IPC spectral index for RQ and RL EMSS AGN subsamples
shows that the mean index of RL sample is flatter than that of the RQ. 
Moreover, as already obtained by Lawson et al. (1992) with the EXOSAT data, 
the RL sample has a 90 per cent confidence contour consistent with zero intrinsic 
dispersion ($i.e$ compatible with an unique spectral index) whereas there is
surely 
an intrinsic dispersion present in the RQ sample. These results support 
a scheme where in radio-loud objects there is an additional X-ray component  
characterized by a single dominant power-law. 

4. Although we have ROSAT data for only 5 RL EMSS AGN, we find that also in the 
ROSAT band the mean index of RL is flatter than that of RQ. This suggests that 
the additional component in RL AGN dominates the X-ray spectra of RL AGN over 
almost two decades of energy ($\sim0.1-10$ keV). 

5. We found no significant correlation between the spectral index $\alpha_x$ 
and other physical parameters such redshift, optical and X-ray luminosity
and the ratio between optical and X-ray luminosities. 
Similar  results were obtained for other X-ray selected sample of AGN

\section*{Acknowledgments}
We thank  R. Della Ceca, S. Molendi, F. Fiore and A. Wolter for useful 
discussions.
PC acknowledges the partial support and hospitality of the Smithsonian 
Astrophysical Observatory (SAO) where part of this work was carried out. 
PC also thanks M. Elvis and B. Wilkes 
for encouragements and useful discussion during his stay at SAO.  
This work has received partial financial support from the Agenzia Spaziale
Italiana (ASI contract:  191/3 AXG)

\vspace{2cm}

\begin{center}
{\bf Figure Captions}
\end{center}

{\bf Figure 1.} Comparison between the spectral slope $\alpha$ determined from 
the source hardness ratio and from a detailed spectral analysis. 

\vspace{3mm}

{\bf Figure 2.} Confidence contours (68, 90 and 99 per cent) for the joint
determination of the energy spectral index and equivalent hydrogen absorbing 
column N$_H$, for six of the eight sources that show significant deviation 
from a simple power-law fit plus Galactic absorption. The vertical line represents
the Galactic N$_H$ value (Stark et al. 1992)

\vspace{3mm}

{\bf Figure 3.} Ratio of the counts in the energy intervals 1.2$-$2.4 keV and 
0.6$-$1.0 keV as a function of the ratio in the intervals 0.6$-$1.0 keV and 
0.2$-$0.4 keV for $\alpha$=0.5, N$_H$=1.2$\times10^{22}$ cm$^{-2}$ ({\it 
crosses}); $\alpha$=0.7, N$_H$=1.2$\times10^{22}$ cm$^{-2}$ ({\it filled
squares}); $\alpha$=0.7, N$_H$=1.5$\times10^{22}$ cm$^{-2}$ ({\it open
squares}) and  $\alpha$=0.9, N$_H$=1.2$\times10^{22}$ cm$^{-2}$ ({\it open
circles}). In all cases U is varied between 0.1 and 0.2 (adapted from 
Fiore et al. 1993).  The points numbered one to four represent the 
EMSS AGN with a best fit value of N$_H$ lower than Galactic value. 
 
\vspace{3mm}

{\bf Figure 4.} Distribution of the energy spectral indices obtained from the analysis
of the ROSAT data. The Radio-loud objects are shaded. 

\vspace{3mm}

{\bf Figure 5.} 90 per cent confidence contours obtained with the 
Maximum-Likelihood analysis for the whole 
EMSS AGN sample with the $Einstein$/IPC data (dotted line) and for the 
subsample of 55 EMSS AGN with the ROSAT/PSPC data (solid line) (see \S 5.2
for more details).

\vspace{3mm}

{\bf Figure 6.} As in Figure 5, for Radio-loud (RL) and Radio-quiet (RQ) AGN
separately. 

\vspace{3mm}

{\bf Figure 7.} ROSAT X-ray spectral indices for Radio-quiet  AGN (open squares) 
and for
Radio-loud AGN (filled squares) versus: (a) Redshift, (b) optical luminosity 
at 2500\AA, (c) X-ray 
broad band (0.1$-$2.4 keV) luminosity and (d) $L_{\rm 2500\AA}/L_x$.

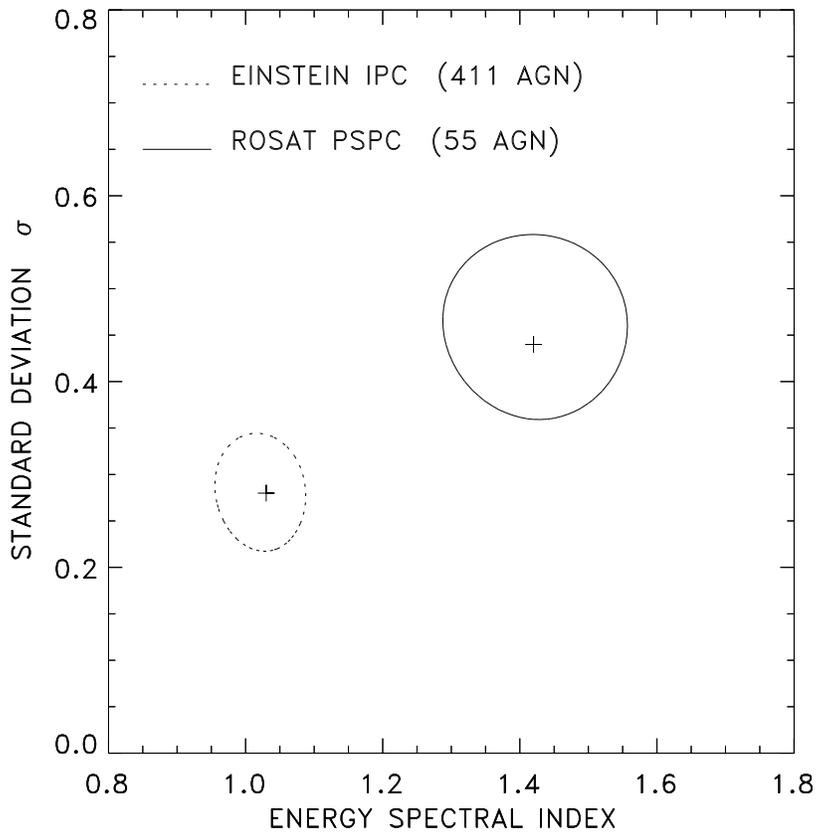
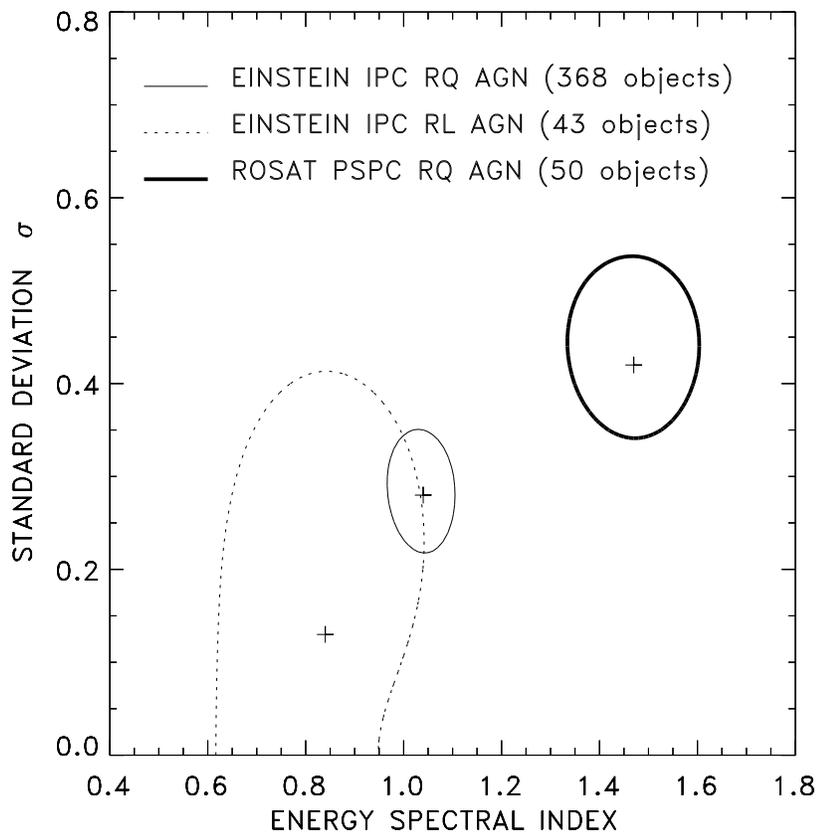
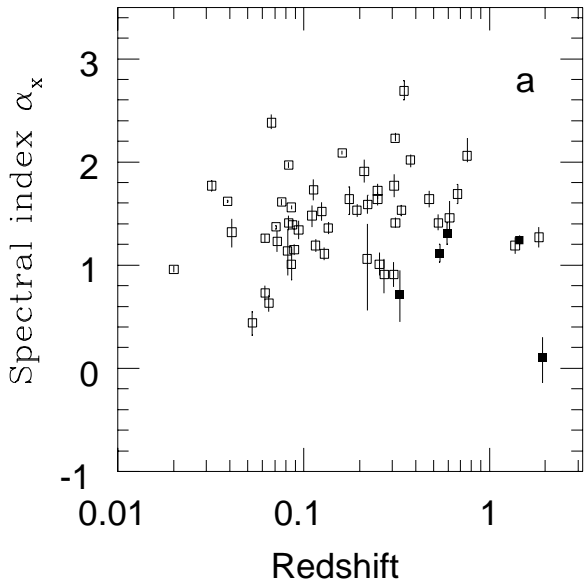
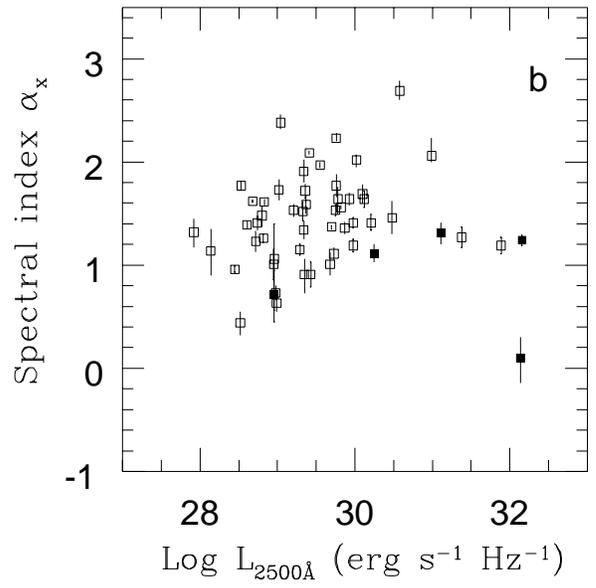
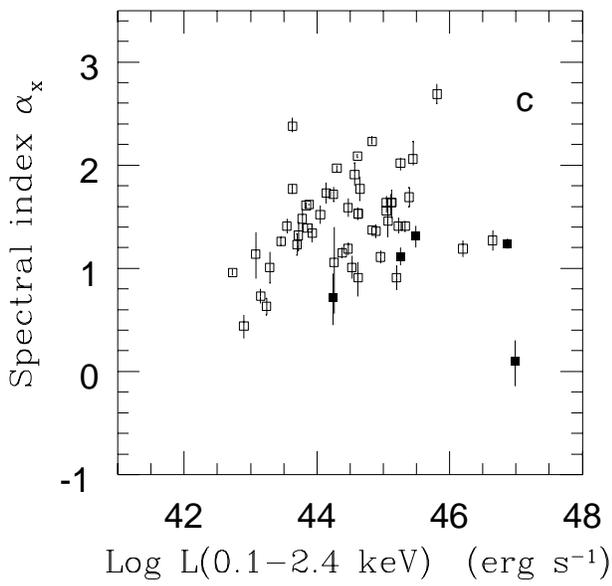
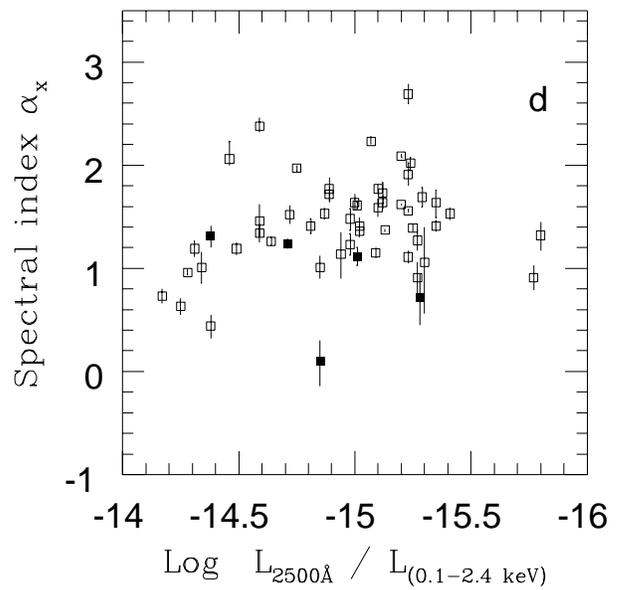
\end{document}